\documentclass{iopart}
\usepackage{iopams}
\usepackage{graphicx}
\usepackage{epstopdf}
\def\bc{\begin{center}}          \def\ec{\end{center}}

\begin{document}

\title{Analytical model of brittle destruction based on hypothesis of scale similarity}

\author{A S Arakcheev and K V Lotov}

\address{Budker Institute of Nuclear Physics, 630090, Novosibirsk, Russia, \\
Novosibirsk State University, 630090, Novosibirsk, Russia}
\ead{asarakcheev@gmail.com}

\begin{abstract}
The size distribution of dust particles in nuclear fusion devices is close to the power function. A function of this kind can be the result of brittle destruction. From the similarity assumption it follows that the size distribution obeys the power law with the exponent between $-4$ and $-1$. The model of destruction has much in common with the fractal theory. The power exponent can be expressed in terms of the fractal dimension. Reasonable assumptions on the shape of fragments concretize the power exponent, and vice versa possible destruction laws can be inferred on the basis of measured size distributions.
\end{abstract}

\pacs{52.40.Hf, 52.25.Vy, 05.45.Df} \maketitle

\section{Introduction}

The dust appears in most of nuclear fusion devices due to
the plasma-wall interaction. The dust particles pose potential
problems to plasma confinement: decrease the plasma temperature,
absorb tritium, etc. Therefore, it is necessary to investigate
dust formation mechanisms and parameters of the dust. The size
distribution is an important characteristic of the dust, as it
determines the dust surface area. The latter in turn determines
the evaporation rate of the dust and the tritium absorption rate.
To our notion, there is no analytical model explaining
experimentally observed dust size distributions.

\begin{figure}[htbp]
\begin{center}
\includegraphics[width=13cm]{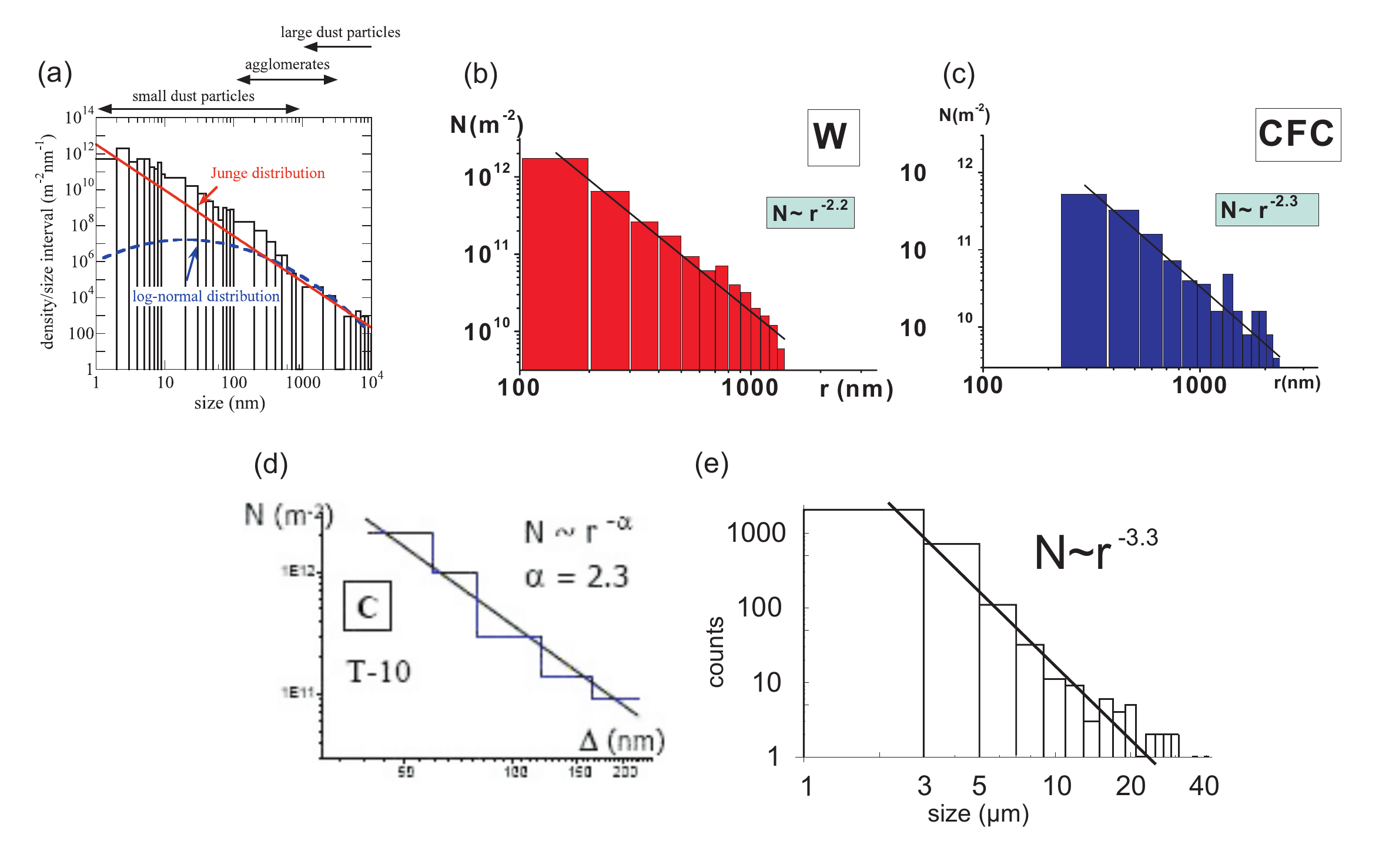}
\end{center}
\caption{(Color online) Dust size distributions measured at (a) stellarator LHD (from \cite{1}), (b,c) QSPA facility (from \cite{10}), (d) tokamak T-10 (from \cite{2}),  and (e) mirror trap GOL-3 (data taken from \cite{13}).}
\label{Power_distribution}
\end{figure}

There are two basic distributions commonly used for approximation
of experimental results: the log-normal distribution and the power
one (the Junge distribution) \cite{24,1,10}. The Junge
distribution was observed in several experiments
\cite{1,10,4,2,13} for tungsten and carbon dust in the size range
from several nanometers to tens of microns
(Fig.~\ref{Power_distribution}). The exponent $(-\alpha)$ of the
power distribution was measured to fall between $-3.3 $ and $-2.1
$. The power distribution was observed both in fusion devices and
in specialized facilities designed for erosion studies.

The power size distribution was also observed for the atmospheric dust \cite{24}. The distribution was explained by the model of particle coagulation \cite {25}. Coagulation results in formation of dust agglomerations which are distributed in size according to the power law. In plasma devices, the power size distribution was observed not only for agglomerations, but also for single dust particles in the size range from several nanometers to several microns \cite{1}. Hence it follows that the coagulation model cannot explain the dust size distribution in the whole range of sizes.

Four basic mechanisms of dust formation are usually distinguished in plasma devices: flaking of redeposited layers, brittle destruction, condensation from the supersaturated vapor, and growth from hydrocarbon molecules \cite{30}. If the dust grows from the gas (condensation from the supersaturated vapor and growth from hydrocarbon molecules), the size distribution differs from the power law \cite {31}. The typical size of the dust formed by flaking is greater than one micron \cite{32}. At smaller sizes, flaking can be thought of as brittle destruction. The brittle destruction is thus the only mechanism that can be responsible for formation of the observed dust. In this paper we show that the brittle destruction can indeed provide the power size distribution.

Brittle destruction was studied theoretically and experimentally \cite{21,8,26,27,28,29}.
The power law for single dust particles was observed in the wide range between two typical sizes: the grain size of materials and interatomic distance. This fact suggests that the law of material fragmentation is independent of the scale. Moreover, the power distribution itself, which intrinsically has no typical size scale, points to the same conclusion. In Section~\ref{matmod} we find which dust size distributions can be realized under the assumption of scale similarity and show that these are the power ones. The exponent of the distribution function can vary in the range from $-4 $ to $-1 $. In Section~\ref{fractals} we show the interrelation between the discussed model and fractal theory and connection of the distribution exponent with the dimension of packing. In Section~\ref{adas} we relate basic features of fragmentation with  distribution functions of fragments for several representative cases. In Section~\ref{discus} we analyze available experimental data and discuss possible implications of measured power exponents.

\section {Mathematical model} \label{matmod}

First we formulate the problem mathematically. The dust particles produced due to the brittle destruction are fragments of a solid body. The body is to be divided into pieces according to some law. Under the assumption of scale similarity, the division law must be independent of the fragment size. Let us mentally remove the pieces from the body one by one in the order of decreasing size and number them sequentially by the index $n$. The size distribution function does not depend on the shape of the initial body only for fragments of the size much smaller than the typical size of the initial body. Hence we assume $n \gg 1$.

Denote the volume of the body remained after removal of  the $n$-th piece by $V_n $, and its surface area by $S_n $. We can write recurrent expressions relating $V_n $, $S_n$, and the characteristic size $r_n $ of the $n$-th fragment:
\begin{equation} \label{2} V_n =V_{n-1}-c_1 r_n^3, \end{equation}
\begin{equation} \label{3} S_n =S_{n-1}+c_2 r_n^2, \end{equation}
where $c_1 $ and $c_2 $ are coefficients that depend on the shape of removed fragments. In what follows we consider these coefficients to be independent of the fragment size and its number due to the scale similarity assumption. The coefficient $c_1$ is positive because the body volume decreases as we remove fragments. The coefficient $c_2 $ can be either positive or negative.

It is necessary to express $r_{n+1}$ in terms of $V_n $ and $S_n$ to close the recurrent scheme. After
removal of many pieces, the size of next fragments is much smaller than typical sizes of the initial body. The structure of the residual body does not depend any more on the shape of the initial body and is determined by the shape of fragments. As $n$ increases, the local structure of the body remains self-similar because the construction law is the same at all scales. The difference is only in the size of structure elements. From the similarity assumption it follows that the size of removed fragments is proportional to the size of these elements. The linear size of the structure element can be determined as the ratio of its volume to the surface area. Under this definition, the result does not depend on the quantity of  elements considered, since both the volume and the surface area are proportional to this quantity. Thus we obtain the required expression:
\begin{equation}\label{4} r_{n+1}=c_3 {V_n} / {S_n}, \end{equation}
where $c_3 $ is a positive coefficient dependent on the shape of removed fragments.

It is convenient to analyze the sequence $q_n = {S_n^3} / {V_n^2} $ instead of the sequences $V_n $, $S_n$, and $r_n $. Its recurrent expression found from (\ref{2}), (\ref{3}) and (\ref{4}) does not contain other variables:
\begin{equation} \label{1} q_{n+1} =q_n \frac {\left (1 + {c_2 c_3^2} / {q_n} \right)^3}
{\left (1 - {c_1 c_3^3} / {q_n} \right)^2}. \end{equation}
This sequence tends to a nonzero constant $q_\infty$ or to infinity. The second case is realized only under the condition
\begin{equation} \label{5} 3 c_2 +2 c_1 c_3> 0. \end{equation}

In the first case, recurrent expressions (\ref{2}) and (\ref{3}) can be simplified for $n \gg 1$:
\begin{equation} \label{8} V_{n+1} =V_n \left (1 -{c_1 c_3^3} / {q_\infty} \right), \end{equation}
\begin{equation} \label{9} S_{n+1} =S_n \left (1 + {c_2 c_3^2} / {q_\infty} \right). \end{equation}
Formula (\ref{1}) in this limit reads as
\begin{equation} \label{10} \left (1- {c_1 c_3^3} / {q_\infty} \right)^2 = \left (1 + {c_2 c_3^2} / {q_\infty} \right)^3. \end{equation}
It follows from formulae (\ref{8}) and (\ref{9}) that the sequences $V_n $ and $S_n $ are geometrical progressions:
\begin{equation}\label{8a} V_{n} \propto \left (1 - {c_1 c_3^3} / {q_ \infty} \right)^n, \end{equation}
\begin{equation}\label{9a} S_{n} \propto \left (1 + {c_2 c_3^2} / {q_ \infty} \right)^n. \end{equation}
Substituting (\ref{8a}) and (\ref{9a}) into (\ref{4}) gives
\begin{equation} r_{n} \propto \left (1 - {c_1 c_3^3} / {q_ \infty} \right)^{n/3}, \end{equation}
where we used expression (\ref{10}) for simplification. Hence it follows the size distribution function of the fragments:
\begin{equation} f(r) = \left | \frac {d n} {d r} \right | \propto \frac {1} {r}. \end{equation}
This result has not been observed in experiments, therefore we focus our attention on the second case.

If the sequence $q_n $ tends to infinity, it is possible to simplify the expression (\ref{1}) in the limit of large $q_n$:
\begin{equation} q_{n+1} \approx q_n \left (1 + \frac {3 c_2 c_3^2 +2 c_1 c_3^3} {q_n} \right)
=q_n + \left (3 c_2 c_3^2 +2 c_1 c_3^3 \right). \end{equation}
For $n \gg 1$, the sequence $q_n $ grows linearly:
\begin{equation}\label{e14} q_n \sim \left (3 c_2 c_3^2 + 2 c_1 c_3^3 \right) n. \end{equation}
The dependence of $V_n $ on $n $ follows from (\ref{2}) and (\ref{e14}):
\begin{equation} {V_{n+1}} / {V_n} =1-{c_1 c_3^3} / {q_n}, \end{equation}
whence we obtain the asymptotic behavior of $V_n $:
\begin{equation} \label{11} V_{n} \propto n^{-\frac {c_1 c_3^3} {3 c_2 c_3^2 +2 c_1 c_3^3}}. \end{equation}
Similarly, the expression for the surface area of the remained body follows from formulae (\ref{3}) and (\ref{e14}):
\begin{equation} \label{12} S_{n} \propto n^{\frac {c_2 c_3^2} {3 c_2 c_3^2 +2 c_1 c_3^3}}. \end{equation}
The substitution of (\ref{11}) and (\ref{12}) into (\ref{4}) gives
\begin{equation} r \propto n^{-\frac {c_2 c_3^2+c_1 c_3^3} {3 c_2 c_3^2 +2 c_1 c_3^3}}. \end{equation}
This dependence yields the size distribution function
\begin{equation} \label{6} f (r) = \left | \frac {dn} {dr} \right | \propto r^{-\alpha }, \qquad
\alpha=3 + \frac {1} {1+ c_1 c_3/c_2}. \end{equation}
From the condition (\ref{5}), we find that $\alpha$ falls into the interval from 1 to 4. All known experimental results are in this interval.

In the case of a $N $-dimensional body, similar calculations give the allowable range for the exponent between $-N-1 $ and $-1 $.

\section{Analogy with fractals}\label{fractals}

The above model of body decomposition resembles an algorithm of fractal construction and can be treated in terms of the fractal theory. There is a relation between the exponent of fragments size distribution and the fractal dimension of packing. The latter is determined as Hausdorff dimension of the residual set \cite{35}.

The above model of fragmentation assumes division of the body into
an infinite number of domains filling the body volume completely.
The same situation is realized at tiling packings, for instance,
at osculatory packings by spheres \cite{33,34}. However, tiling
packings are generally non-self-similar. This fact makes the
involved mathematics significantly more complex. Therefore we
derive the key relation with a self-similar fractal like
Sierpinski carpet \cite{35, 3} which is not a classical object of
the packings theory.

For fractals of this kind, a number of equal-size fragments is
removed at every construction stage. The process is characterized
by two parameters: the ratio of linear sizes of fragments removed
at consequent stages ($k_1 >1$), and the ratio of numbers of
fragments removed at consequent stages ($k_2 >1$). Denote the
number of stage by $i $, then we can write for the fragment number
$n$ and the fragment size $r$
\begin{equation}n=1+k_2+k_2^2+...+k_2^i=\frac{k_2^{i+1}-1}{k_2-1},\end{equation}
\begin{equation}r\propto k_1^{-i},\end{equation}
and find the relation between them for $i\gg 1$:
\begin{equation}n\propto r^{-\log_{k_1}{k_2}}.\end{equation}

The $\log_{k_1}{k_2} $ equals to fractal dimension $D$ \cite{35,3}. Thus the fragment size distribution function is
\begin{equation}\label{e23} f\left(r\right)=\left|\frac{d n}{d r}\right|\propto r^{-1-D}, \qquad
\alpha = 1+D.\end{equation}
The same expression was proved for some non-self-similar fractals, particularly for osculatory packings \cite{37,38}. Possibly it is a general law for every decomposition described in these terms.

The fractal dimension of the residual set in three-dimensional
space evidently falls between $0$ and $3$. The interval for
possible values of $\alpha$ thus naturally follows from the
dimension of space. However, the result of Section~\ref{matmod} is
more general, since fractional-dimensional nature is not assumed there and the
exponent is related to the shape of fragments.

\begin{figure}[]
\begin{center}
\includegraphics[width=8.5cm]{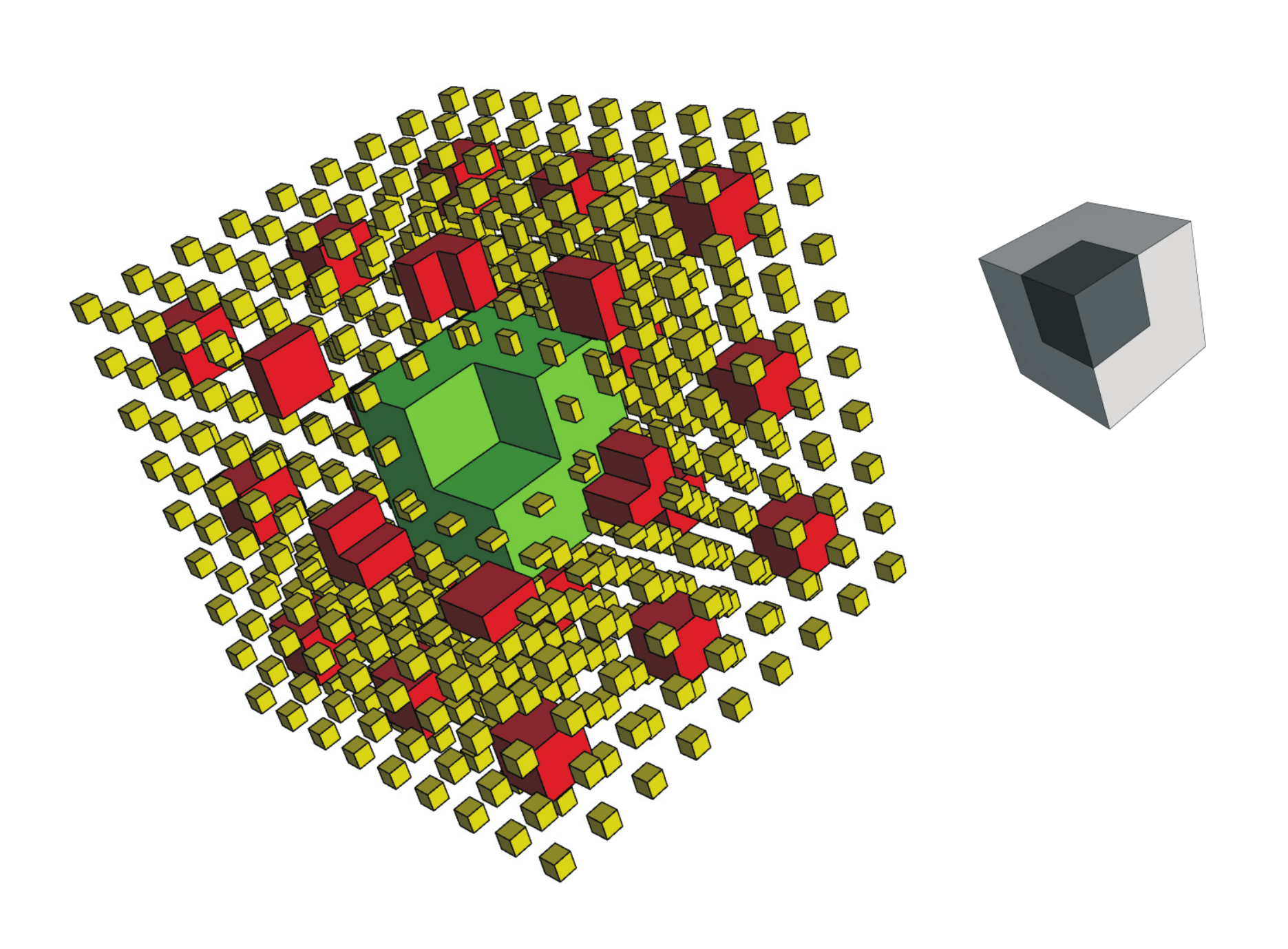}
\end{center}
\caption{(Color online) Three generations of fragments (marked by different colors) of Sierpinski cube having a segment cut away. The inset shows which segment is cut away.}
\label{Cube}
\end{figure}

Let us illustrate the geometry of fragmentation for different
values of $\alpha$. The case of large $\alpha$ may be presented by
Sierpinski cube (Fig.~\ref{Cube}), i.e. the three-dimensional analogue of the
Sierpinski carpet \cite{35}, for which $\alpha \approx 3.97 $. The special feature of the Sierpinski cube is that it is ``spongy''. 

More rigorously, the combination of coefficients in the expression (\ref{6})
can be interpreted in this way:
\begin{equation} \label{100} \frac{c_1 c_3}{c_2}={-\frac{\Delta V_n}{\Delta S_n}}\biggm/{\frac{V_n}{S_n}},\end{equation}
where $\Delta V_n=V_{n+1}-V_n$ and $\Delta S_n=S_{n+1}-S_n$. The difference
$\Delta V_n$ always equals the volume of the $n$-th fragment
$V_n^*$. The quantity $\Delta S_n$ can differ from the area  of the $n$-th
fragment $S_n^*$ due to partial coincidence of the fragment surface with the
surface of the fragmented body. In the Sierpinski cube, the fragments do not
touch each other, so that $\Delta S_n = S_n^*$, and $c_2$ has the maximum (positive) value possible for fragments of this shape and size. Consequently,
the expression (\ref{100}) takes the form
\begin{equation} \label{101} \frac{c_1 c_3}{c_2}={-\frac{V_n^*}{S_n^*}}\biggm/{\frac{V_n}{S_n}}.\end{equation}
The surface area of the removed fragment is
the same as the surface contained in one cell of the cube, but the
volume of the residual in the cell is 26 times bigger. So the value of
(\ref{101}) is small ($\approx$\,1/26), and $\alpha$ in (\ref{6}) is close
to $4$.

\begin{figure}[]
\begin{center}
\includegraphics[width=8.5cm]{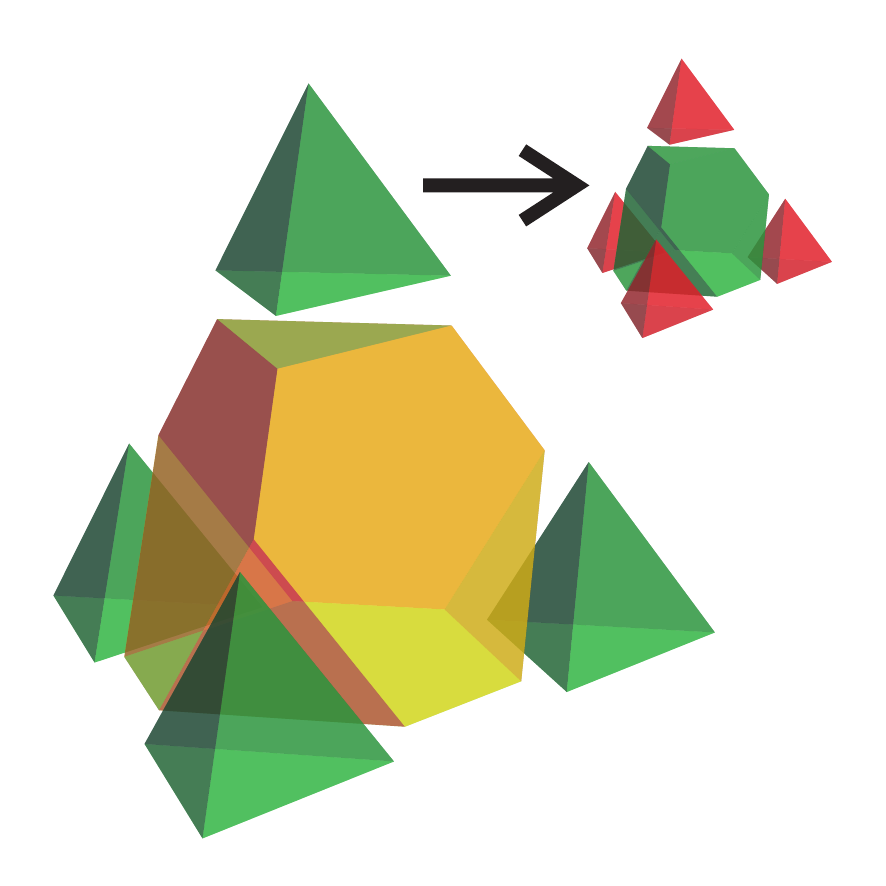}
\end{center}
\caption{(Color online) The scheme of fragmentation for the modified Sierpinski tetrahedron with the fractal dimension~$1.4$.}
\label{Tetrehedron}
\end{figure}

The three-dimensional analogue of Sierpinski triangle
\cite{35} is a tetrahedron from which the central octahedron is removed
so that four tetrahedrons of the twice shorter edge remain.
The exponent in distribution (\ref{e23}) for it is $-3 $ sharp. This is the separating case corresponding to $c_2=0 $, which means the surface area does not change in the fractal construction process.
If the removed fragment is larger and four remaining tetrahedrons are smaller (Fig.~\ref{Tetrehedron}), then $k_1> 2 $, $k_2=4 $, $c_2 <0$, and thus the exponent is smaller in absolute value. The limit of very small remaining tetrahedrons corresponds to $\alpha = 1 $, which is another limiting value.

Note that the discussed fractals do not fully conform to
the introduced model of destruction: the coefficient $c_3 $ oscillates as a function of the
fragment number. Therefore, it is necessary to take an average value of $c_3 $.

In Ref.~\cite{2}, a fractal structure of separate dust particles falling on a substrate was detected. The fractal dimension of particles was measured to be $2.2 \pm 0.2$, which is close to the coefficient $\alpha \approx 2.3 \pm 0.1$ for the same dust. This coincidence conflicts with relation (\ref{e23}) that follows from the very basics of the fractal theory. Thus the used method of measuring the fractal dimension is open to question, although the observation of a fractal structure is important by itself.

\section{Additional assumptions}\label{adas}

Experiments show significant diversity in values of $\alpha$ which, apparently, is caused by diverse fragmentation laws in different experimental setups. It is difficult to uniquely derive the fragmentation law from the exponent. Therefore we analyze several reasonable fragmentation mechanisms, calculate the parameter $\alpha$ for them, and correlate basic features of fragmentation with observable distribution functions of fragments.

First, we assume shape regularity of fragments. In the general case, $ \Delta V_n=V_n^* $ and
$ \Delta S_n \leq S_n^* $. Whence it follows
\begin{equation}\label{13}\left|\frac{\Delta V_n}{\Delta S_n}\right|\geq\frac{V_n^*}{S_n^*}.\end{equation}
Substitution of (\ref{13}) into (\ref{101}) gives
\begin{equation}\label{14}\left|\frac{c_1 c_3}{c_2}\right|\geq\left|\frac{V_n^*}{S_n^*}\biggm/{\frac{V_n}{S_n}}\right|.\end{equation}
If the shape of each fragment is more regular than the shape of the residual, then the
right-hand side of inequality (\ref{14}) is much greater than
unity, as is the left-hand side. Consequently, the exponent in the
size distribution function (\ref{6}) differs from $-3 $ by a small parameter. In the case of
$N$-dimensional body, similar calculations give the
exponent close to $-N $.

Second, we take into account the experimentally observed shape of
fragments. In Ref.~\cite{1}, single small dust particles were
photographed, and their shapes are spherical. Shapes of large
($\gtrsim 1 \, \mu$m) particles are irregular. Filling of space
with spheres was studied in connection with osculatory packing
problem \cite{33,34}. The key feature of the osculatory packing is
that the sphere of maximum possible radius is inscribed into the
residual at each step. The osculatory packing in three-dimensional
space gives $\alpha \approx 3.47$ \cite{36}. The osculatory
packing by disks in two-dimensional space corresponds to $\alpha
\approx 2.3$.

Third, we use energy considerations. The brittle destruction is
more energetically efficient than evaporation, but it also needs
energy for surface tension. If this energy is a factor, then
fragmentation must happen with minimal surface formation. Any
packing by spheres is energy ineffective since spheres can touch
neighbours only in one point, and no surface is shared between
fragments. An example of energy effective fragmentation is the
modified Sierpinski tetrahedron for which the surface area of
the residual decreases and the total surface area of all fragments is limited.

There is a compromise solution that meets both sphericity and energy considerations.
At the modified Sierpinski tetrahedron, the parameter $k_1$ is free. We
can vary it to make fragments as spherical as possible. We take the ratio
${V_n^*}/{S_n^*} $ as a criterion of sphericity; this combination is maximal for a ball. The maximum for the modified Sierpinski tetrahedron corresponds to $\alpha \approx 2.4$ (Fig.~\ref{Tetrehedron}).

\section{Discussion}\label{discus}

Several references \cite{10,2} give experimental results for $\alpha$ in the range from $2.2$ to $2.3$ with an accuracy of $0.1$. Close values are given by osculatory packing by disks ($\alpha \approx 2.3$) and modified Sierpinski tetrahedron ($\alpha \approx 2.4$). This suggests that material destruction at described facilities is either energy-abundant two-dimensional or energy-scarce three-dimensional. The former variant seems more realistic since the penetration depth for particle energies involved is shorter than the smallest size of observed dust particles.

There are experiments showing greater values of the coefficient $\alpha$ \cite{13,Shiratani}.  In Ref.\,\cite{13}, not only size distribution was measured in the interval from 2\,$\mu$m to 40\,$\mu$m, but also the depth of erosion $\sim 10\,\mu$m in a single shot. The strong erosion appears due to hot electrons that deeply penetrate into the target and explode it volumetrically. These results could demonstrate the transition from the two-dimensional fragmentation to the three-dimensional one. Indeed, for small particles $\alpha\approx 3.3$, which is clearly distinguished from the value of $2.3$ in \cite{10,2}. Unfortunately, for particles larger than 10\,$\mu$m the statistics is poor and no quantitative analysis is possible.

The exponent close to $-3.47$ corresponding to three-dimensional osculatory packing by spheres is observed for several non-fusion objects \cite{39,40}. First, the size distribution of interstellar grains has the power in the interval from $-3.6$ to $-3.3$ for variety of materials. Second, brittle destructed materials (coal mine dust, crushed lead glass) has the fractal dimension corresponding to $\alpha \approx 3.5$. Of course, the measured dimension is the dimension of fragment surface rather than that of the residual, but they are equal in this case.

The observed distributions can be modified by collisions of fragments. If collisions round corners of fragments like sea waves polish pebbles through their contact, then the resulting law of fragmentation will approach the osculatory packing by spheres. Thus, the exponent close to $-3.5$ may bear witness to long-term collisional evolution of fragments.

There are several factors not included into the model which can result in difference between theoretical predictions and measurements. Counting the dust particles and measuring their size are performed after impaction against the substrate. The particles can coagulate or change shape on the way to the substrate or on the impact, can overlap each other on the substrate, etc. Therefore, to extract maximum information from the developed model, a great care is to be taken to minimize evolution of the dust on the way to the detector.

\ack

The authors greatly appreciate stimulating discussions with S.Arakcheev, A.Burdakov, I.Garkusha, M.Shiratani, A.Shoshin, D.Skovorodin,and A.Sudnikov.
This work is supported by by RF President's grant NSh-7792.2010.2, Russian Ministry of Education grants RNP-2.1.1/3983 and 14.740.11.0053, and RFBR grant 11-02-00563.

\end{document}